# ONLINE SENSOR TESTING THROUGH SUPERPOSITION OF ENCODED STIMULUS

*Norbert Dumas[1], Zhou Xu[1], Kostas Georgopoulos[1], John Bunyan[2], Andrew Richardson[1]*

[1]Centre for Microsystems, Lancaster University, Lancaster, LA1 4YR, UK
[2]QinetiQ, St Andrews Road, Malvern, Worcs, WR14 3PS, UK

**ABSTRACT**

Online monitoring remains an important requirement for a range of microsystems. The solution based on the injection of an actuating test stimulus into the bias structure of active devices holds great potential. This paper presents an improved solution that aims to remove the measurand-induced signal from the sensor output. It involves encoding the test stimulus and using a covariance algorithm to reject the signal that does not contain the code. The trade-off between the sine wave rejection ratio of the technique and the test time response is studied and, in the case of a MEMS accelerometer, it is demonstrated that the rejection is higher than 14dB for a test time of about 0.7s. Furthermore, the accuracy of the test signal can be evaluated to guarantee the integrity of the online test output.

## 1. INTRODUCTION

The online test technique where a stimulus is superimposed through the sensor bias has been presented in [1] for an accelerometer and in [2] for a conductance sensor. In order to actuate the device without affecting the operational output, the stimulus frequency is chosen to be outside the operating bandwidth but within the physical bandwidth. The measurand-induced signal is removed by a filter assuming that its frequency is different to that of the test signal. However, this assumption cannot always be guaranteed because the measurand is most of the time not controlled and therefore can be at the same frequency as the test signal. Consequently, the test output can fluctuate and cause the test to produce erroneous results.

To solve this issue, it is proposed to modulate the test sine wave stimulus by a pseudo-random bit code sequence (Maximum Length Sequence) generated by a LFSR. The stimulus response is processed through a covariance algorithm, which measures the amplitude of the response, and a correlation algorithm, which measures how much the response, is affected by a perturbing signal. The sensor is reported faulty only if the covariance output is in range and the correlation process ensures that the test response is not perturbed by the measurand signal. A pseudo random sequence is utilized in order to avoid a perturbation signal (measurand) exhibiting a similar waveform to that of the code. The underlying principle is similar to spread spectrum communication [3] where the quasi constant power spectral density of the code at low frequencies reduces the probability of interference. Moreover, it is well established that test techniques based on LFSR offer good noise rejection [4]. It should be noted that the use of the covariance technique differs from other typical maximum length sequence test techniques that calculate the complete impulse response by using the convolution method. In this paper, where online configuration is addressed, the impulse response cannot be calculated because the bandwidth available for test is narrow.

In section 2, the case study accelerometer is presented. The principle of the technique is described in section 3. In section 4, the technique is theoretically evaluated. The implementation considerations are discussed in section 5.

## 2. CASE STUDY SENSOR

The capacitive MEMS accelerometer taken as a case study has the same parameters as in [1]. The maximum acceleration is limited to 15g and the resolution is around 1mg over a 100Hz bandwidth. It acts like a second order low pass system that is critically damped to avoid ringing of the output. The overall physical bandwidth of the sensor is 1.3kHz.

The test acceleration can be electro statically induced through additional comb fingers or the bias. Although the the conversion of electrical stimulus into equivalent acceleration is in effect non-linear, we only consider extremely small movement of the proof mass and hence use a linearised model for the transfer function. In this paper the test stimulus is directly applied as acceleration ($A_t$) with amplitude of 50mg (corresponding to a 0.9V applied to the comb finger). The relationship between the output voltage and the input acceleration $A$ is:

$$V_{out} = G \cdot k_c \cdot \frac{1}{M \cdot s^2 + D \cdot s + K} \cdot M \cdot A \qquad (1)$$

where $A=A_m+A_t$, $K$, $M$ and $D$ are the spring, mass and damping coefficients, $G$ is the amplifier gain, $k_c$ is a coefficient of proportionality between the displacement of the mass and the output capacitance variation.

## 3. PRINCIPLE

### 3.1. Superposition without encoded stimulus

In the basic superposition scheme presented in [1], the stimulus is a sine wave signal (see Figure 1). Its frequency (1kHz for the case study) falls into the physical bandwidth of the sensor (1.3kHz for the case study) but outside the operational bandwidth limited by the operational filter (100Hz for the case study). With a 50mg test stimulus and an operational filter attenuation of at least -40dB at the test frequency, the test induced signal is lower than the noise level at the output $V_{out}$. Consequently, the test stimulus has no impact on the normal operation of the sensor.

The test filter enables measurement of the sensor response to the stimulus. For the case study it is a high pass filter that efficiently attenuates the signals below 100Hz at the sensor output. The amplitude of the test output is related to the sensor sensitivity at the test frequency. If its absolute amplitude variation exceeds a threshold, the sensor is reported as faulty. It should be noted that other types of test than this sensitivity test can be used online with the superposition techniques, like for example the differential test that checks if the sensor symmetry is broken. These kinds of techniques potentially offer better fault coverage, however in this study a sensitivity test is chosen to illustrate the effectiveness of the technique.

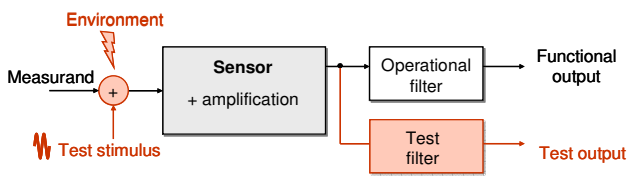

**Figure 1: Basic superposition scheme**

The issue encountered with the scheme shown in Figure 1 appears when the physical signal has a component around the test frequency. In this case the test output fluctuates and an erronous decision of the sensor status can be reported by the test electronics. For example, if the case-study accelerometer measures the low frequency acceleration for a navigation application, the perturbing signal can come from vibration induced by the environment. The fluctuation of the test output is illustrated in Figure 2 when the perturbing signal is as large as the test signal in terms of equivalent input acceleration. To evaluate the sensitivity of the output to perturbation we define the sensitivity to perturbation (SP) and the rejection (Rej) as:

$$SP = \frac{1}{Rej} = \frac{|\Delta V_{test}|/V_{ref}}{|A_p|/|A_t|} \quad (2)$$

where $V_{ref}$ is the amplitude of the test signal without perturbation and $\Delta V_{test}$ is the maximal error $V_{test} - V_{ref}$. Without encoding the sine wave stimulus the rejection of a perturbation that has its power concentrated around the test frequency (not filtered by the sensor or the test filter) is equal to 0dB.

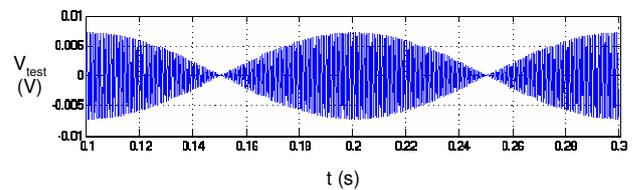

**Figure 2 Test output fluctuation with a 50mg equivalent test acceleration at 1kHz and a 50mg vibration at 1003kHz**

### 3.2. Superposition of an encoded stimulus

The proposed solution to encode the test stimulus is presented in Figure 3. The code generated by the LFSR is a sequence of -1 and +1 modulating a sine wave carrier at the test frequency $f_t$ (1kHz). Without the low pass filter $LP_2$ placed after the LFSR the resulting test stimulus would have 90% of its total power concentrated between $f_t-f_c$ and $f_t+f_c$, $f_c$ being the code bit rate. Therefore $f_c$ is limited by the bandwidth available for the test. The low pass filter ($LP_2$) is included in the design to remove part of the test stimulus power that could perturb the operational output. The signal at the output of the sensor is filtered and then demodulated to retrieve the low frequency code. Special care must be taken to ensure that the signal to be demodulated is in phase with the carrier signal.

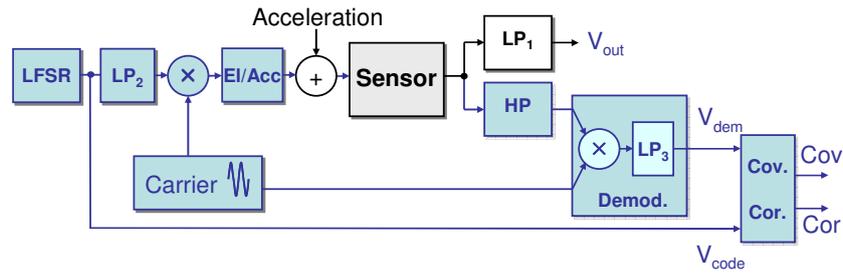

**Figure 3: Schematic of the proposed solution implemented on an accelerometer**

The demodulated signal ($V_{demod}$) is converted into the digital domain and its covariance and correlation with one sequence of the input code calculated. The test scalar value obtained (*Cov* and *Cor*) are both used to take a decision on the sensor status. The result can then be compared to a reference value in order to generate a flag that can either indicate a dysfunction in the sensor or transmitted to a higher system level as a parameter that reflects how much the sensor is degraded.

Covariance (*Cov*) and correlation (*Cor*) are calculated using a full code sequence in the following manner:

$$Cov = \text{cov}(V_{dem}, V_{code}) = E(V_{dem} V_{code}) - E(V_{dem}) E(V_{code}) \quad (3)$$

$$Cor = cor(V_{dem}, V_{code}) = \frac{\text{cov}(V_{dem}, V_{code})}{\sigma_{dem} \cdot \sigma_{code}} \quad (4)$$

where $\sigma_x$ is the standard deviation of $V_x$ and E() is the average function defined as:

$$E(V_x) = \sum_{n=1}^{N} \frac{V_x[n]}{N} \quad (5)$$

where *N* is the number of samples taken over a sequence.

*Cov* is proportional to the stimulus-induced signal whilst any irrelevant signals are attenuated. *Cor* is close to 1 if the demodulated signal matches exactly the code and decreases when other perturbing signals become superposed. If the *Cov* result varies significantly with respect to a reference value, it means that the sensitivity of the sensor has varied or that a strong perturbation is present and not sufficiently attenuated. This latter cause is detected by checking that the *Cor* value has decreased significantly. Decision on the sensor status is made only if the *Cor* value is higher than a certain threshold.

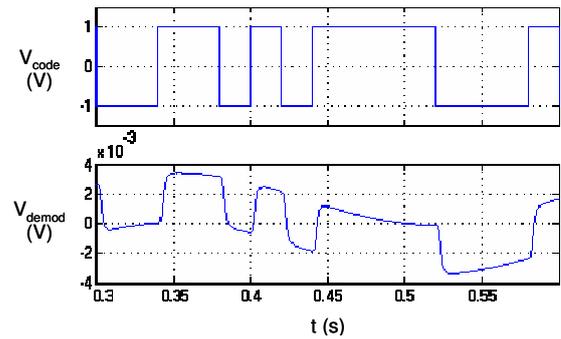

**Figure 4: Above - The 8 bits input code @50Hz. Below- The demodulated test output affected by a 1003Hz/50mg sine wave acceleration.**

Figure 4 illustrates by simulation the perturbation of the demodulated signal by a perturbation around the test frequency. Despite the fact that the fluctuation of the demodulated signal is as large as the output code amplitude, the variation of the covariance output is only 5%, resulting in 26.6dB of rejection according to the equation (2). It clearly shows that the perturbing signal is attenuated.

### 4. EVALUATION OF THE TECHNIQUE

#### 4.1. Expression of the rejection

To evaluate the performance of the technique, a sine wave signal has been considered to model the perturbation effects. For the accelerometer under study, the signal can come from environmental vibration coupled into the test frequency range. The demodulated signal $V_{dem}$ is therefore the sum of the signal induced by the encoded test signal (amplitude=$M_A$) and a signal generated by a sine wave (amplitude=$k \cdot M_A$, where *k* is the amplitude ratio). We will make the approximation that these signals are not attenuated by the filters. In practise we have seen that the high frequency components of the code are attenuated. However this is a minor approximation because the primary component of power in the code is concentrated at low frequency (between DC and $f_c$). We will therefore express $V_{dem}$ as:

$$V_{dem} = \alpha \cdot M_A \cdot [V_{code} + k\cos(2\Pi ft + \varphi)] \quad (6)$$

where $\alpha$ is a coefficient proportional to the sensor sensitivity at the test frequency $f_t$; $f$ and $\varphi$ are the frequency and phase of the down-shifted sine wave (i.e. $f = |f_p - f_t|$ where $f_p$ is the frequency of the perturbation).

As a result of the distributive property of the covariance, the output $Cov$ is expressed as :

$$Cov = \alpha M [\text{cov}(V_{code}, V_{code}) + k\,\text{cov}(V_{code}, \cos(2\Pi ft + \varphi))] \quad (7)$$

The first term of the equation is what we want to measure ($Cov_{ref}$) and the second is the error due to the perturbation. Since the covariance of the code signal with itself is equal to 1, dividing the relative error by the amplitude ratio $k$ according to equation 2, we obtain:

$$SP = \text{cov}(V_{code}, \cos(2 \cdot \Pi \cdot f + \varphi)) \quad (8)$$

Hence, the sensitivity to perturbation is equal to the covariance of the code signal with a sine wave. Similarities between this expression and the expression of the coefficient of the discrete Fourier transform of $V_{code}$ ($C_f$) have been found. The complex coefficients of the discrete Fourier transform of $V_{code}$ are given by:

$$C_k = \sum_{n=1}^{N} V_{code}[n] \cdot \cos\left(\frac{2\Pi}{N}(k-1)(n-1)\right) \\ - i\sum_{n=1}^{N} V_{code}[n] \cdot \sin\left(\frac{2\Pi}{N}(k-1)(n-1)\right) \quad (9)$$

where $k = 1, 2, \ldots, N$.

This above equation can be written differently by using the covariance. Using the definition of the covariance given by the equation (3), we can express the covariance of the code with a cosine and a sine wave of frequency $f$ taking discrete value $f = (k-1)/(N\tau_s)$, $k = 1, 2, \ldots, N$ as follow:

$$\text{cov}(V_{code}, \cos(2 \cdot \Pi \cdot \frac{k-1}{N\tau_s} \cdot t)) = \\ \frac{1}{N}\sum_{n=1}^{N} V_{code}[n] \cdot \cos\left(2\Pi \frac{k-1}{N}(n-1)\right) - E(V_{code}) \cdot E(\cos) \quad (10)$$

$$\text{cov}(V_{code}, \sin(2 \cdot \Pi \cdot \frac{k-1}{N\tau_s} \cdot t)) = \\ \frac{1}{N}\sum_{n=1}^{N} V_{code}[n] \cdot \sin\left(2\Pi \frac{k-1}{N}(n-1)\right) - E(V_{code}) \cdot E(\sin) \quad (11)$$

To simplify, we can assume that the second term of these two equations is very small. The first reason is due to a property of the code generated by LFSR. The sequence contains only one more '1' (+1) than '0' (-1). As a consequence the average value $E(V_{code})$ is very small. The second reason is that the average value of a cosine or sine wave is zero over a finite number of periods and is very close to zero for a large number of periods. Therefore $E(\cos)$ and $E(\sin)$ tend towards zero for increasing frequency $f$. Replacing the equations (10) and (11) without their second term into equation (9), we obtain the following approximation:

$$C_k \approx N\,\text{cov}(V_{code}, e^{-i\frac{2\Pi}{N}(k-1)(n-1)}) \quad (12)$$

From the equation (8), we can deduce that:

$$SP \approx \frac{1}{N}\text{Re}(e^{i\varphi}\overline{C}_f) \quad (13)$$

This approximation assumes that $f$ is sufficiently high ($f > 1/\tau$ where $\tau$ is the time length of the code). Expression (13) is maximal for $\varphi = \arg(C_f)$, i.e. the worst case, is:

$$SP_{max} \approx \frac{|C_f|}{N} \quad (14)$$

### 4.2. Analysis of the rejection

Figure 5 shows the plot of $SP_{max}$ with respect to frequency $f$ that corresponds to a perturbation signal of frequency $f_p = f \pm f_t$. This has been generated using the definition of covariance presented in equation (3) (Cov. curve) and from equation (14) (FFT curve). As expected a difference is only observed at low frequencies. From this graph we can see that the worse rejection value is around 6.7dB.

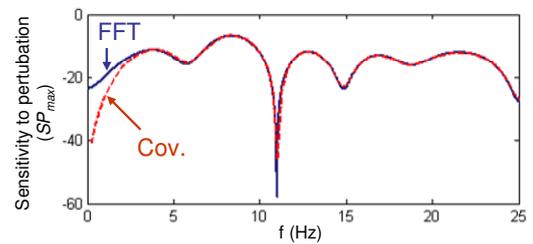

**Figure 5:** $SP_{max}$ **with respect to frequency of the perturbation (down-shifted).** $N_b$=15bits@50Hz.

Pseudo random codes have well known spectral properties. The overall envelope of the power spectrum has the shape of a sine cardinal function [3]:

$$envelope(PSD(f)) = \frac{1}{f_c}\left(\frac{\sin(\Pi f / f_{clock})}{\Pi f / f_{clock}}\right) \quad (15)$$

The power spectral density of the code can be deduced from the discrete Fourier transform coefficients ($C_f$):

$$PSD(f) = \frac{|C_f|^2 \tau_s}{N} \quad (16)$$

The envelope of equation 15 represents an average of equation 16. The envelope reaches its maximum for $f = 0$ and be considered flat for low frequencies. Therefore a good approximation of the Fourier transform coefficients ($C_f$) at low frequencies is:

$$|C_f|_{around\,f=0} \approx \sqrt{\frac{N}{f_c \tau_s}} \quad (17)$$

From equation (17) and (14), we can deduce that:

$$SP_{max} \approx \sqrt{\frac{1}{N_b}} \quad (18)$$

This expression represents the Root Mean Square (RMS) value of the sensitivity to perturbation over the flat low frequency band of the code spectrum. From this expression, the sensitivity of the output test signal to a perturbation depends solely on the number of bits in the sequence and decreases by -3dB/octave with respect to the $N_b$, figure. However the overall test time and bit rate poses a limiting factor on the number of bits that can be employed by the technique. Thus, the highest acceptable bit rate with no effects on the normal operation of the sensor is 100Hz. Given that the specified maximal test time is 1second, it follows that the maximum number of bits is 63@100Hz (6th order LFSR). The resulting RMS sensitivity to perturbation is -17.8dB, the maximal sensitivity to perturbation is around -14dB and the test time is 0.63s.

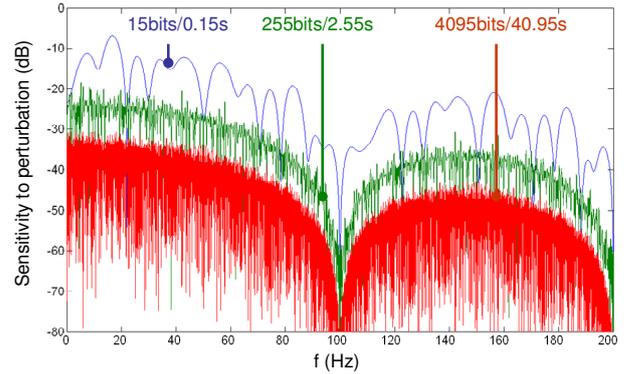

**Figure 6: Evolution of the sensitivity to perturbation with respect to the sequence length ($f = 0$ corresponds to $f_p = f_t = 1$kHz)**

### 4.3. Use of correlation

At this stage, despite the rejection achieved by the proposed solution, the probability of fluctuation is not zero. For instance, consider that the maximal acceleration around the test frequency is 15g and that the required equivalent input resolution of the test signal is 1mg, i.e. the accuracy is 2%. In that case the worst rejection ratio that would be acceptable is 83.5dB. This value is significantly larger than the best rejection ratio that can be achieved. However, it is not very likely that a perturbing vibration would continuously exceed 50mg. Assuming that this type of perturbation are non-stationary, the following approach is proposed. The aim is to detect vibrations that introduce corruption at the test output (*Cov*).

The detection of the perturbation level is performed by correlating the code and the demodulated signal. Contrary to covariance, this mathematical operation is not sensitive to the absolute value of the stimulus and perturbation-induced signals but it is sensitive to the amplitude ratio (*k*) between the two. When there is no perturbation ($k = 0$), the correlation result is equal to 1 (Figure 7). Consequently, increasing values of *k* generate a decreasing trend in the correlation coefficients. Considering that the rejection ratio is equal to or higher than 14dB, if the cross correlation drop is lower than 3% of its reference value, the amplitude ratio *k* is guaranteed to be lower than 0.4 (figure 1). This value of *k* is related to the error signal by the following formula deduced from equation 2:

$$\frac{|\Delta Cov|}{Cov_{ref}} = kSP \quad (19)$$

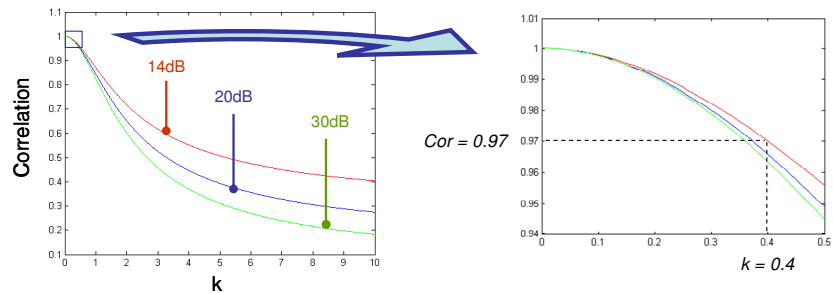

**Figure 7:** Correlation output with respect to *k* for rejection ratio = 14dB, 20dB and 30dB

For the worst case (*SP* = -14dB), this corresponds to an accuracy of 8%. If the correlation is lower than the 97% threshold, the accuracy is no longer guaranteed. In such case, the system does not make a decision on the sensor status and is not updated. Another test code sequence has to be launched. Possibly, the system may increase dynamically the length of the code. This will increase the rejection ratio and then decrease the probability that the test signal is corrupted.

## 5. IMPLEMENTATION

The results have been verified by simulating the complete architecture, illustrated in figure 1, with Simulink. Concerning the analogue implementation of the architecture, the main elements which are required are three filters, two mixers and one analogue-to-digital converter. The sine wave carrier can also be replaced by a square wave signal. The benefit is to replace the two mixers with analogue switches. For the digital implementation, the LFSR, the covariance and correlation are simple functions and could be implemented using a micro controller. In this study the sampling frequency needed is only 800Hz, corresponding to 8 samples per bit.

## 6. CONCLUSION

In this paper, a novel architecture for online sensor testing has been presented. Extensive analysis has shown that it can significantly improve the rejection of a sine wave perturbation and offer a way to evaluate the accuracy of the test signal. As a consequence, the test structure is guaranteed to produce good results regarding the sensor status. The method can be applied to any linear sensor provided that is a means of electrically inducing the stimulus and that a certain part of the sensor bandwidth can be reserved for test. Future work is focused on implementing the proposed technique on the accelerometer used as a demonstrator in this paper.

Stimulus injection is achieved through the accelerometer bias and the non-linearity of the injection technique is studied. The additional test cost for the implementation is expected to be low.

## 7. ACKNOWLEDGEMENT

This work is being supported by the EPSRC Innovative Electronics manufacturing Research Centre (IeMRC) through the I-Health project (Integrated Health Monitoring of MNT based Heterogeneous Systems (SP/06/03/07) in the UK and the European FP6 Network of Excellence in Design for Micro & Nano Manufacture (PATENT-DfMM), Project no. 507255.